\documentclass[11pt]{article}
\usepackage{geometry}                
\usepackage{amsmath}
\usepackage{graphicx}
\usepackage{dcolumn}
\usepackage{bm}
\usepackage{wrapfig}
\usepackage{epsfig}
\usepackage{breqn}
\usepackage{amssymb}
\usepackage{subfigure}
\usepackage[usenames,dvipsnames]{pstricks}
\usepackage{epsfig}
\usepackage{pst-grad} 
\usepackage{pst-plot} 
 \usepackage[usenames,dvipsnames]{pstricks}
 \usepackage{epsfig}
 \usepackage{pst-grad} 
 \usepackage{pst-plot} 
 \usepackage{float}

\usepackage[all]{xy}

\begin{document}

\title{Stochastic Eternal Inflation in a Bianchi Type I Universe}
\author{Ikjyot Singh Kohli \\ isk@mathstat.yorku.ca \\ York University - Department of Mathematics and Statistics
\and Michael C. Haslam \\ mchaslam@mathstat.yorku.ca \\ York University - Department of Mathematics and Statistics}

\date{December 9, 2015}

\maketitle 

\begin{abstract}
The phenomenon of stochastic eternal inflation is studied for a chaotic inflation potential in a Bianchi Type I spacetime background. After deriving the appropriate stochastic Klein-Gordon equation, we give details on the conditions for eternal inflation. It is shown that for eternal inflation to occur, the amount of anisotropy must be small. In fact, it is shown that eternal inflation will only take place if the shear anisotropy variables take on values within a small region of the interior of the Kasner circle. We  then  calculate the probability of eternal inflation occurring based on techniques from stochastic calculus. 
\end{abstract}

\section{Introduction}
The theory of inflation arose out of trying to explain the so-called flatness and horizon problems in cosmology \cite{hervik}. However, the vast majority of non-eternal and eternal inflationary models assume from the outset that the background spacetime is spatially homogeneous, flat, and isotropic, and are therefore considered in a $k=0$ Friedmann-Lema\^{i}tre-Robertson-Walker (FLRW) spacetime \cite{elliscosmo}. Stochastic eternal inflation has received a considerable amount of attention over the past few years as it is one of main motivations behind the multiverse \cite{elliscosmo}. In this paper, we study the dynamics of stochastic eternal inflation in the context of a Bianchi Type I spacetime, which is spatially homogeneous, but anisotropic, and hence, is more general than the standard $k=0$ FLRW cosmology. It is of considerable interest to consider whether stochastic eternal inflation could happen in a more general background spacetime as this would allow the possibility of the existence of a multiverse to arise from more generic conditions, in addition to solving the so-called ``fine-tuning'' problem in a more general sense.  To the best of the authors' knowledge, such a study has not been presented in the scientific literature.

Non-eternal inflationary dynamics in the context of Bianchi Type I cosmological models have been studied a number of times in the literature. Barrow and Turner \cite{1981Natur.292...35B} showed that a universe with moderate anisotropy will undergo inflation and will rapidly be isotropized. Rothman and Marsden \cite{1985PhLB..159..256R} analyzed in detail inflationary dynamics in Bianchi Type I cosmologies.  Jensen and Stein-Schabes \cite{1986PhRvD..34..931J} studied the effects of anisotropic cosmologies on inflation in significant detail. Moss and Sahni \cite{1986PhLB..178..159M} studied the effect of anisotropies on the chaotic inflationary for a general class of Bianchi cosmologies. Rothman and Ellis \cite{1986PhLB..180...19R} showed that inflation takes place in all Bianchi models and that anisotropy does not suppress inflation. Belinsky and Khalatnikov \cite{belkhatgrossman} studied homogeneous cosmological models of Bianchi I and Friedmann type in the presence of a massive scalar field. They showed that the majority of solutions undergo an inflationary stage.  Gr{\o}n \cite{1990Ap&SS.173..191G} considered inflationary Bianchi Type I models with shear, bulk, and nonlinear viscosity. Kitada and Maeda \cite{1992PhRvD..45.1416K} proved a cosmic no-hair theorem for Bianchi models in power-law inflation. Parnovskij \cite{1993ZhETF.103..337P} analyzed the evolution of several Bianchi Type models in the presence of a scalar field. It was shown that certain Bianchi models reduce to the Bianchi Type I model in this case. Burd and Barrow \cite{Burd:1988ss} considered cosmological models with a scalar field with an exponential potential. They investigated the inflationary nature of a $n+1$-dimensional Friedmann universe and some $3+1$0-dimensional cosmological models as well. Aguirregabiria, Feinstein, and Ib{\'a}{\~n}ez \cite{1993PhRvD..48.4662A} obtained a general exact solution of the Einstein field equations for Bianchi type I universes filled with an exponential-potential scalar field and also studied their dynamics. Ib{\'a}{\~n}ez, van den Hoogen, and Coley \cite{1995PhRvD..51..928I} studied whether homogeneous cosmological models containing a self-interacting scalar field with an exponential potential isotropize. Harko \cite{1996AcPhH...3..115H} studied the dynamics of a Bianchi type I universe filled with a scalar field, and also discussed conditions under which inflation can occur in such models. Chimento \cite{1998CQGra..15..965C} found the exact general solution for cosmological models arising from the interaction of the gravitational field with two scalar fields in both flat FLRW and the locally rotationally symmetric Bianchi Type I spacetimes filled with an exponential potential. Byland and Scialom \cite{1998PhRvD..57.6065B} studied the Einstein-Klein-Gordon equations for a convex positive potential in a Bianchi Type I, a Bianchi Type III, and a Kantowski-Sachs universe. After analyzing the inherent properties of the system of differential equations, they studied the asymptotic behaviors of the solutions and their stability for an exponential potential. Chakraborty and Paul \cite{2001PhRvD..64l7502C} studied the cosmic no-hair theorem for anisotropic Bianchi models which admit an inflationary solution with a scalar field. It was found that the form of the potential does not affect the evolution in the inflationary era. Chimento \cite{2002PhRvD..65f3517C} showed that inflation can be obtained from nonaccelerated scenarios by a symmetry transformation in Bianchi Type I spacetimes. Bali and Jain \cite{2002Prama..59....1B}  investigated a Bianchi Type I inflationary universe in the presence of massless scalar field with a flat potential, and discussed the inflationary scenario in detail. Fay and Luminet \cite{2004CQGra..21.1849F} studied in detail different conditions leading to the isotropization of a Bianchi Type I model. Folomeev \cite{2007IJMPD..16.1845F} studied the dynamics of a Bianchi Type I model with two interacting scalar fields. Gumrukcuoglu, Contaldi, and Peloso \cite{Gumrukcuoglu:2007bx} studied inflationary perturbations in a Bianchi Type I model and studied their imprint on the cosmic microwave background.  Pitrou, Pereira, and Uzan \cite{2008JCAP...04..004P} investigated the predictions of an inflationary phase starting from a Bianchi Type I universe. Beyer and Escobar \cite{2013CQGra..30s5020B} considered the dynamics of Bianchi A scalar field models which undergo inflation, and in particular, studied the ``graceful'' exit problem. 

Stochastic eternal inflation has also been studied a number of times in different contexts.  Miji\'{c} \cite{mijic1} analyzed the boundary conditions of the dynamics of inflation as a relaxation random process and gave a simple proof for the existence of eternal inflation. Salopek and Bond \cite{salobond} studied nonlinear effects of the metric and scalar fields in the context of stochastic inflation. Linde, Linde, and Mezhlumian \cite{lindelindemezh} considered chaotic inflation in theories with effective potentials that behave either as $\phi^n$ or as $e^{\alpha \phi}$. They also performed computer simulations of stochastic processes in the inflationary universe. Linde and Linde  \cite{lindelinde} investigated the global structure of an inflationary universe both by analytical methods and by computer simulations of stochastic processes in the early universe. Susperregi and Mazumdar \cite{susmazum} considered an exponential inflation potential and showed that this theory predicts a uniform distribution for the Planck mass at the end of inflation, for the entire ensemble of universes that undergo stochastic inflation. Vanchurin, Vilenkin, and Winitzki \cite{vanvilwin} investigated methods of inflationary cosmology based on the Fokker-Planck equation of stochastic inflation and direct simulation of inflationary spacetime. Winitzki \cite{winitzki} explored the fractal geometry of spacetime that results from stochastic eternal inflation. Kunze \cite{kunze} considered chaotic inflation on the brane in the context of stochastic inflation. Winitzki \cite{winitzki2} described some issues regarding the time-parameterization dependence in stochastic descriptions of eternal inflation. Gratton and Turok \cite{grattonturok} investigated a simple model of $\lambda \phi^4$ inflation with the goal of analyzing the continuous revitalization of the inflationary process in some regions. Li and Wang \cite{liwang1} used a stochastic approach to investigate a measure for slow-roll eternal inflation. Tom\`{a}s G\`{a}lvez Ghersi, Geshnizjani, Piazza, and Shandera \cite{tomas} used stochastic eternal inflation to analyze the connection between the Einstein equations and the thermodynamic equations. Qiu and Saridakis \cite{qiusarid} used stochastic quantum fluctuations through a phenomenological, Langevin analysis studying whether they can affect entropic inflation eternality. Harlow, Shenker, Stanford, and Susskind \cite{harlowshenker} described a discrete stochastic model of eternal inflation that shares many of the most important features of the corresponding continuum theory. Vanchurin \cite{vanchurin1} developed a dynamical systems approach to model inflation dynamics. Feng, Li, and Saridakis \cite{fenglisaridakis} investigated conditions under which phantom inflation is prevented from being eternal. Recently, Kohli and Haslam \cite{2015CQGra..32g5001S} showed that the stochastic eternal inflation scenario in the context of a spatially flat FLRW cosmology leads to strong singularities, thus eliminating the possibility of inflation being eternal in such models.

In this paper, we consider the effects of adding a stochastic forcing term in the form of Gaussian white noise to the right-hand-side of the Klein-Gordon equation as is done in \cite{grattonturok} and \cite{2015CQGra..32g5001S}. Following \cite{elliscosmo}, we note that the classical dynamics of the inflaton dictates that the inflaton always rolls down its potential. However, quantum fluctuations can also drive the inflaton uphill, which causes inflation to last longer in some regions, which, in turn, enlarges the volume of the region. In some regions, the inflaton will remain high enough up the potential hill to maintain acceleration. This is a stochastic scenario that is typically described as \emph{eternal} inflation which is one of the main motivating ideas behind the multiverse concept. 
As discussed in \cite{grattonturok}, the purpose of this stochastic term is to describe the scalar field fluctuations. 

Note that, throughout, we assume geometrized units, $c=8\pi G=1$, so that all dynamical quantities have units of powers of length. 

\section{The Dynamical Equations}
In this paper, we are considering a Bianchi Type I universe. We will in this section, employ the orthonormal frame formalism \cite{ellismac} to the Einstein field equations to obtain a dynamical system of equations. Following \cite{ellis}, we consider a group-invariant orthonormal frame $\{\mathbf{n}, \mathbf{e}_{\alpha}\}$, where $\mathbf{n}$ is the unit normal to the group orbits. Further, since $\mathbf{n}$ is tangent to a hypersurface-orthogonal congruence of geodesics, all dynamical variables depend only on time, $t$. The Einstein field equations take the form
\begin{eqnarray}
\label{eq:raych1}
\dot{H} &=& -H^2 - \frac{2}{3}\sigma^2 - \frac{1}{6} \left(\mu + 3p\right), \\
\label{eq:shear1}
\dot{\sigma}_{\alpha \beta} &=& -3H \sigma_{\alpha \beta}, \\
\label{eq:fried1}
\mu &=& 3H^2 - \sigma^2,
\end{eqnarray}
where $H$ is the Hubble parameter, $\sigma_{\alpha \beta}$ is the shear tensor, and $\mu$ and $p$ are the energy density and pressure of the matter content in the cosmological model respectively. 

As mentioned in the introduction, our universe model contains a scalar field, which is minimally coupled and has the Lagrangian density
\begin{equation}
\mathcal{L} = -\sqrt{g} \left[\frac{1}{2} \nabla_{a} \phi \nabla_{a}\phi + V(\phi)\right],
\end{equation}
which implies that the energy-momentum tensor takes the form
\begin{equation}
T^{ab} = \nabla^a \phi \nabla^b \phi - \left[\frac{1}{2} \nabla_c \phi \nabla^c \phi + V(\phi)\right]g^{ab}.
\end{equation}

Following \cite{elliscosmo}, since we are working within a $3+1$ formalism, such that we assume that $\nabla_a \phi$ is timelike, and is normal to the space like surfaces $\phi(x^{a}) = const$, the energy-momentum tensor has the algebraic form of a perfect fluid, with
\begin{equation}
\mu = \frac{1}{2}\dot{\phi}^2 + V(\phi), \quad p = \frac{1}{2}\dot{\phi}^2 - V(\phi).
\end{equation}

The Klein-Gordon equation which is implied from the divergence-free property of $T_{ab}$ above then becomes
\begin{equation}
\ddot{\phi} + 3H \dot{\phi} + V'(\phi) = 0.
\end{equation}
To model the quantum fluctuations in the eternal inflation scenario, we add a stochastic noise forcing term $\eta(t)$ on the right-hand-side of this equation, so that it takes the form
\begin{equation}
\label{eq:kg1}
\ddot{\phi} + 3H \dot{\phi} + V'(\phi) = H^{5/2}\eta(t).
\end{equation}

Further, we also note that since the shear tensor $\sigma_{\alpha \beta}$ is diagonal and trace-free, we define
\begin{equation}
\sigma_{+} = \frac{1}{2} \left(\sigma_{22} + \sigma_{33}\right), \quad \sigma_{-} = \frac{1}{2\sqrt{3}} \left(\sigma_{22} - \sigma_{33}\right), 
\end{equation}
such that
\begin{equation}
\sigma^2 \equiv \frac{1}{2}\sigma_{\alpha \beta}\sigma^{\alpha \beta} = 3 \left(\sigma_{+}^2 + \sigma_{-}^2\right).
\end{equation}

In this paper, we will specifically consider the case of a chaotic inflation potential, $V(\phi) = \frac{1}{2}m^2 \phi^2$, and will also make the slow-roll approximation, where $V(\phi) > \dot{\phi}^2$. Within the slow-roll approximation, the Friedmann equation \eqref{eq:fried1}, takes the form
\begin{equation}
V(\phi) = 3H^2 - \sigma^2.
\end{equation}
We will also employ expansion-normalized variables, defining, 
\begin{equation}
\label{eq:expan1}
\Sigma^2 \equiv \frac{\sigma^2}{3H^2}, \quad X \equiv \sqrt{\frac{V(\phi)}{3}} \frac{1}{H},
\end{equation}
such that
\begin{equation}
X^2 + \Sigma^2 = 1.
\end{equation}

Under the slow-roll approximation, the Klein-Gordon equation \eqref{eq:kg1} becomes
\begin{equation}
3H\dot{\phi} + V'(\phi) = H^{5/2}\eta(t).
\end{equation}
Letting $V(\phi) = \frac{1}{2}m^2\phi^2$, and employing the expansion-normalized variables in Eq. \eqref{eq:expan1}, one obtains
\begin{equation}
X' -(1+q)X + m_{0}^2 X = \mathcal{N}(\tau),
\end{equation}
where we have defined
\begin{equation}
\label{eq:newdefs}
\mathcal{N}(\tau) = \frac{\eta}{3\sqrt{6H}}, \quad m_{0} = \frac{m}{\sqrt{3} H}. 
\end{equation}
Note that $m_{0}$ is the expansion-normalized dimensionless inflaton mass, which in our units has dimensions of $H$ \cite{waldbook}.

Further, $q$ is the deceleration parameter which is found from Raychaudhuri's equation \eqref{eq:raych1} to be
\begin{equation}
\label{eq:qdefzero}
q = 2\Sigma^2 - X^2.
\end{equation}

Finally, the expansion-normalized shear propagation equations take the form
\begin{equation}
\Sigma_{\pm}' = \Sigma_{\pm}(q-2).
\end{equation}

Note that, in the expansion-normalized variables, differentiation is denoted by a primed symbol, and is with respect to a dimensionless time variable $\tau$, which is defined as \cite{ellis} $dt/d\tau = 1/H$.

The Gaussian noise term $\mathcal{N}(\tau)$ deserves some explanation, which we now provide and is based on arguments given in \cite{2015CQGra..32g5001S}. We will actually make use of the fact that the noise term is formally defined as the time derivative of the Wiener process $W(\tau)$, that is,
\begin{equation}
\label{eq:wien1}
W'(\tau) \equiv \mathcal{N}(\tau), \quad W(0) = 0.
\end{equation}
It should be noted that this definition defines the Gaussian noise term as the \emph{formal} derivative of the Wiener process. As we show below, the Wiener process is actually nowhere differentiable, so one has a set of stochastic differential equations as opposed to ordinary differential equations.
As noted in \cite{grattonturok}, $\mathcal{N}(\tau)$ also satisfies the autocorrelation relation
\begin{equation}
\langle \mathcal{N}(\tau) \mathcal{N}(\tau') \rangle = \delta(\tau - \tau').
\end{equation}
Further, it is important to realize that $W$ is nowhere differentiable. In fact, for completeness, following \cite{wiersema} we now state some properties of $W$ based on the so-called L\'{e}vy characterization:
\begin{enumerate}
\item The path of $W$ is continuous and starts at 0,
\item $W$ is a martingale and $[dW(\tau)]^2 = d\tau$,
\item The increment of $W$ over time period $[s,\tau]$ is normally distributed, with mean $0$ and variance $(\tau-s)$,
\item The increments of $W$ over non-overlapping time periods are independent.
\end{enumerate}
Note that, we will not go into extensive detail about martingales. The interested reader is asked to consult \cite{karatzas} for more details on martingales and their properties. Based on the arguments provided in \cite{wiersema}, we will now show that $W$ is not anywhere differentiable, that is, not $C^{1}$ anywhere. Consider a time interval of length $\Delta \tau = 1/n$ starting at $\tau$. We will define the rate of change over an infinitesimal time interval $[\tau, \tau + \Delta \tau]$ is
\begin{equation}
X_{n} \equiv \frac{W(\tau + \Delta \tau) - W(\tau)}{\Delta \tau} = \frac{W(\tau + \frac{1}{n} - W(\tau)}{\frac{1}{n}} = n \left[W \left(\tau + \frac{1}{n}\right) - W(\tau)\right].
\end{equation}
Therefore, $X_{n}$ is normally distributed with expectation value $0$, variance $n$, and of course, standard deviation, $\sqrt{n}$. It is clear then that $X_{n}$ has the same probability distribution as $\sqrt{n}Z$, where $Z$ is the standard normal distribution. To analyze the $C^{1}$ properties, we must see what happens to $X_{n}$ as $\Delta \tau \to 0$, in other words as $n \to \infty$. For any $k > 0$, let $X_{n} = \sqrt{n} Z$, then
\begin{equation}
\mathbb{P} \left[ \left\| X_{n} \right\| > k\right] = \mathbb{P} \left[  \left\| Z \right\| > \frac{k}{\sqrt{n}}\right].
\end{equation}
Clearly, as $n \to \infty$, $k / \sqrt{n} \to 0$, so we have that
\begin{equation}
 \mathbb{P} \left[  \left\| Z \right\| > \frac{k}{\sqrt{n}}\right] \to \mathbb{P} \left[ \left\| Z \right\| > 0 \right] = 1.
\end{equation}
The point is that we can choose $k$ to be arbitrarily large, so that the rate of change at time $\tau$ is not finite, and therefore, $W$ is not differentiable at $\tau$. Since $\tau$ is arbitrary, one concludes that $W$ is \emph{nowhere differentiable}. 

In summary, the dynamical system describing the dynamics of our cosmological model is given by the following set of equations
\begin{eqnarray}
\label{eq:Xp}
X' &=& \mathcal{N}(\tau) + X \left(1+q-m_0^2\right), \\
\label{eq:Sp}
\Sigma_{\pm}' &=& \Sigma_{\pm}(q-2), 
\end{eqnarray}
where
$q = 2\Sigma^2 - X^2$, such that
\begin{equation}
\label{eq:fcon1}
\Sigma^2 + X^2 = 1,
\end{equation}
where $\Sigma^2 = \Sigma_{+}^2 + \Sigma_{-}^2$.

\section{The Conditions for Eternal Inflation}
We will now derive conditions for the Bianchi Type I model considered in this paper to exhibit eternal inflation. In the non-eternal, slow-roll inflation scenario, one defines a slow-roll parameter, $\epsilon$ as:
\begin{equation}
\epsilon \equiv -\frac{H'}{H}.
\end{equation}
Comparing with the general definition for the deceleration parameter $q$ \cite{ellis},
\begin{equation}
H' = -(1+q)H,
\end{equation}
we see that slow-roll inflation will take place whenever
\begin{equation}
\label{eq:eps1}
\epsilon < 1 \Rightarrow q < 0.
\end{equation}
However, because we are seeking conditions for \emph{eternal} inflation, this condition may not be sufficient.  As discussed in \cite{Vachaspati:2003de} and \cite{Winitzki:2001fc}, for inflation to be eternal, the background spacetime requires frequent violations of the Null Energy Condition. One can typically take fluctuations of the energy-momentum tensor of the scalar field as a possible source of such violations. In the situation where quantum fluctuations dominate slow-roll evolution, the energy-momentum tensor fluctuations are large enough to create Null Energy condition violations. Thus, in this section,  we shall derive conditions where the quantum fluctuations dominate the slow-roll evolution.

To derive the conditions for eternal inflation, we first note that a stochastic differential equation has the general form
\begin{equation}
\label{eq:dX0}
dX = a(X,t) dt + b(X,t) dW,
\end{equation}
where $a(X,t)$ is known as the drift coefficient which effects the deterministic evolution of the system, while $b(X,t)$ is known as the dispersion coefficient which introduces random/stochastic fluctuations into the system. Using Eq. \eqref{eq:wien1}, we can write Eq. \eqref{eq:Xp} as
\begin{equation}
\label{eq:dX1}
dX = \left[ \left(1+q-m_0^2\right) X\right]d\tau + dW.
\end{equation}
Comparing these two equations, we see that the dispersion coefficient is unity, while the drift coefficient is given by $\left(1+q-m_0^2\right) X$. For eternal inflation to occur, the quantum fluctuations which are represented by the stochastic forcing term in Eq. \eqref{eq:dX1} must dominate over the deterministic slow-roll term \cite{Winitzki:2001fc}. In a stochastic differential equation of the form  Eq. \eqref{eq:dX0}, the dynamics of $X$ will be dominated by stochastic `jumps' if $b(X,t) > a(X,t)$. Therefore, to have eternal inflation, we require that
\begin{equation}
\label{eq:einfcond1}
 \left(1+q-m_0^2\right) X < 1,
\end{equation}
where because of the Friedmann equation \eqref{eq:fcon1}, we have that $-1 \leq X \leq 1$. The restriction given by \eqref{eq:einfcond1} will further constrain \eqref{eq:eps1} as we will now see. The only way to satisfy \eqref{eq:einfcond1} is to take
\begin{equation}
\label{qcond1}
-2 < q < 0, \quad -1 \leq X < 0, \quad 0 < m_{0} < \sqrt{\frac{-1+X+qX}{X}},
\end{equation}
or
\begin{equation}
\label{qcond2}
-2 < q < 0, \quad 0 \leq X \leq 1,
\end{equation}
where, of course, $m_{0} > 0$.
Therefore, as can be seen from \eqref{qcond1} and \eqref{qcond2}, the condition \eqref{eq:eps1} is not sufficient for eternal inflation. Taking $-2 < q <0$ in combination with the conditions in  \eqref{qcond1} and \eqref{qcond2} will ensure that eternal inflation takes place. 

One can also go a step further by writing according to Eq. \eqref{eq:fcon1} $X^2 = 1 - \Sigma^2 = 1 - \Sigma_{+}^2 - \Sigma_{-}^2$. Making this substitution in \eqref{qcond1} and \eqref{qcond2}, we obtain the following restrictions on the anisotropy of the model:
\begin{equation}
\label{eq:sigrestr}
-\frac{1}{\sqrt{3}} < \Sigma_{+} < \frac{1}{\sqrt{3}}, \quad -\frac{\sqrt{1-\Sigma_{+}^2}}{\sqrt{3}} < \Sigma_{-} < \frac{\sqrt{1-\Sigma_{+}^2}}{\sqrt{3}}.
\end{equation}
An equivalent expression can also be obtained by interchanging $\Sigma_{+}$ and $\Sigma_{-}$ in \eqref{eq:sigrestr}. 

\begin{figure}[h]
\centering
\includegraphics[scale=0.8]{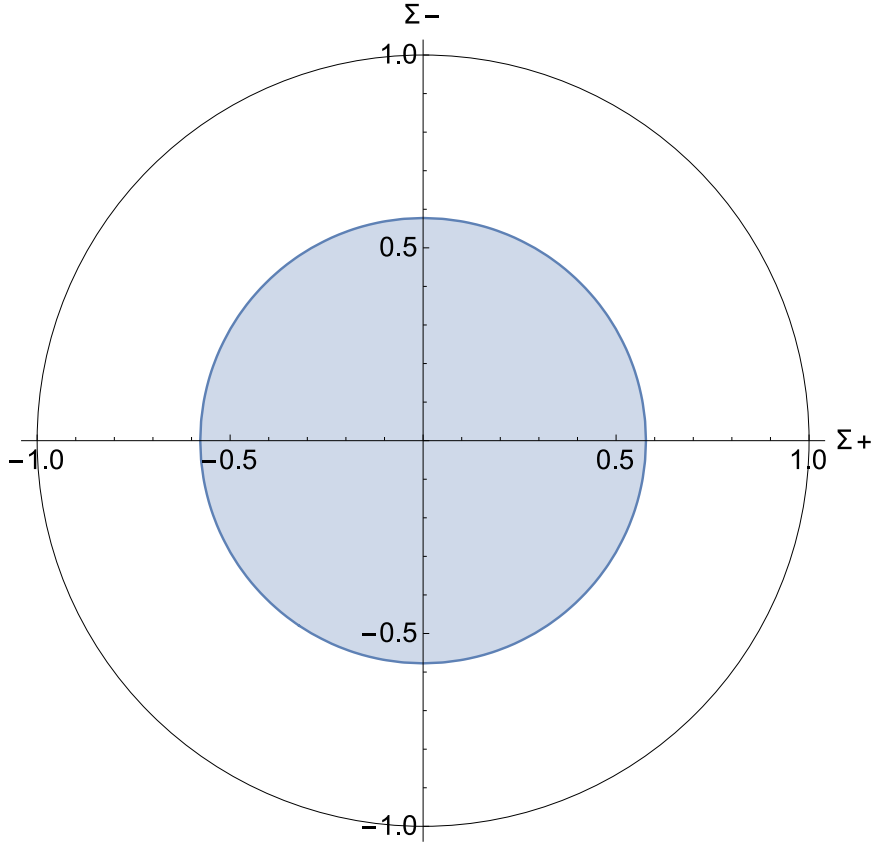}
\caption{A plot of the Kasner circle $\Sigma_{+}^2 + \Sigma_{-}^2 = 1$. For eternal inflation to occur, the anisotropy variables must take on values $(\Sigma_{+}, \Sigma_{-})$ within the shaded region.}
\label{fig:fig3}
\end{figure}

One sees that the anisotropy required for eternal inflation to take place must be contained in a  small region within the Kasner circle, $\Sigma_{+}^2 + \Sigma_{-}^2 = 1$. We have displayed the regions of anisotropy that allow for eternal inflation in Figure \ref{fig:fig3}, based on the conditions \eqref{eq:sigrestr}. These conditions support the claim made in \cite{1986PhLB..180...19R}, where inflation will only occur if there is not a significant amount of anisotropy to begin with.  However, we believe this is the first proof using expansion-normalized variables based on an orthonormal frame approach. 

\section{The Probability of Eternal Inflation}
We now wish to calculate the probability of eternal inflation occurring. We first note that Eqs. \eqref{eq:Xp}-\eqref{eq:Sp} upon using the definition in Eq. \eqref{eq:wien1} take the form
\begin{eqnarray}
dX &=& \left[ \left(1+q-m_0^2\right) X\right]d\tau + dW, \\
d\Sigma_{\pm} &=& \Sigma_{\pm} \left(q-2\right)d\tau,
\end{eqnarray}
where
$q = 2\Sigma^2 - X^2$, such that
\begin{equation}
\label{eq:fried3}
\Sigma^2 + X^2 = 1,
\end{equation}
where $\Sigma^2 = \Sigma_{+}^2 + \Sigma_{-}^2$.

Because of the symmetry of the Bianchi Type I model, from the Friedmann equation, we will write $\Sigma^2 = 1 - X^2$, which implies that $q = 2-3X^2$. So, the inflaton dynamics is governed by a single stochastic differential equation
\begin{equation}
\label{eq:stoch1}
dX = \left[ \left(1+q-m_0^2\right) X\right]d\tau + dW,
\end{equation}
where
\begin{equation}
\label{eq:stochq}
q = 2-3X^2.
\end{equation}
In fact, it is convenient to substitute $q$ into Eq. \eqref{eq:stoch1}, and write
\begin{equation}
\label{eq:stoch2}
dX = \left[X\left(3-m_o^2\right)-3X^3\right]d\tau + dW.
\end{equation}
Because of the highly nonlinear drift term in Eq. \eqref{eq:stoch2}, there is no closed-form solution to Eq. \eqref{eq:stoch2} \cite{iacus}. In fact, what is often of interest in applications of stochastic differential equations is the expected value of the process $X$, which in this case, also has no closed-form expression. Another interesting thing to note is that the deceleration parameter $q$ is now a random variable, as can be seen from Eq. \eqref{eq:stochq}, where $X(\tau)$ is given by  Eq. \eqref{eq:stoch2} as
\begin{eqnarray}
X(\tau) &=& X(0) + \int_{0}^{\tau} \left[ X(s)\left(3-m_o^2\right)-3X(s)^3\right]ds + \int_{0}^{\tau}dW(s) \nonumber \\
 &=& X(0) + \int_{0}^{\tau} \left[X(s)\left(3-m_o^2\right)-3X(s)^3\right]ds + W(\tau).
\end{eqnarray}

Further, recall from the previous section that the condition for inflation to take place is given by the conditions \eqref{qcond1} and \eqref{qcond2}. From Eq. \eqref{eq:stochq}, we see that inflation will occur when
\begin{equation}
\sqrt{\frac{2}{3}} < X \leq 1 \Rightarrow \sqrt{\frac{2}{3}} < \left[X(0) + \int_{0}^{\tau} \left[X(s)\left(3-m_o^2\right)-3X(s)^3\right]ds + W(\tau)\right] \leq 1,
\end{equation}
or
\begin{equation}
-1 \leq X < -\sqrt{\frac{2}{3}} \Rightarrow -1 \leq \left[X(0) + \int_{0}^{\tau} \left[X(s)\left(3-m_o^2\right)-3X(s)^3\right]ds + W(\tau)\right] < -\sqrt{\frac{2}{3}},
\end{equation}
where in this latter equation, we have the condition on $m_{0}$ as described in \eqref{qcond1}.

We are now in a position to calculate the stationary probability distribution. The stationary distribution is of great interest with respect to eternal inflation, since the stationary distribution describes the distribution of $X(\tau)$ over a long period of time. In addition, it is well-known that  inflation should last for more than 60 e-folds \cite{elliscosmo}. We proceed as follows. Following \cite{iacus}, we first compute the scale measure, $s(X)$, as
\begin{eqnarray}
s(X) &=& \exp \left[-2 \int_{X_{0}}^{X}  y\left(3-m_o^2\right)-3y^3 dy \right] \nonumber \\
&=& \exp \left[-2 \left(\frac{3X^2}{2} - \frac{m_o^2 X^2}{2} - \frac{3X^4}{4} - \frac{3X_0^2}{2} + \frac{m_0^2X_{0}^2}{2} + \frac{3X_{0}^4}{4}\right)\right].
\end{eqnarray}
The speed measure, $m(X)$, is then computed as
\begin{eqnarray}
m(X) &=& \frac{1}{s\left(X\right)} \nonumber \\
&=&\exp \left[2 \left(\frac{3X^2}{2} - \frac{m_o^2 X^2}{2} - \frac{3X^4}{4} - \frac{3X_0^2}{2} + \frac{m_0^2X_{0}^2}{2} + \frac{3X_{0}^4}{4}\right)\right].
\end{eqnarray}
The stationary density, $\pi(X)$, is then given by
\begin{eqnarray}
\label{eq:piX}
\pi(X) &=& \frac{m(X)}{\int_{-1}^{1} m(X) dX} \nonumber \\
&=&\frac{e^{-\frac{1}{2} X^2 \left(2 m_0^2+3 X^2-6\right)}}{\int_{-1}^{1} e^{-\frac{1}{2} X^2 \left(2 m_0^2+3 X^2-6\right)} \, dX}
\end{eqnarray}
The motivation behind choosing the limits of integration as equal to plus or minus unity for the stationary density calculation in Eq. \eqref{eq:piX} is the Friedmann equation, Eq. \eqref{eq:fried3}, which tells us that $X = 0$ when $\Sigma^2 = 1$, and  $X = \pm 1$ when $\Sigma^2 = 0$.

It is perhaps of some interest to calculate this probability of eternal inflation occurring using some values for $m_{0}$. Following \cite{1982PhLB..108..389L}, \cite{padma1}, and \cite{dine1}, one can take $H \approx 10^{12} \mbox{ GeV}$ ($10^{-7}$ in natural units), and $m \approx 10^{12} - 10^{13} \mbox{ GeV}$ ($10^{-7}-10^{-6}$ in natural units). We shall therefore calculate the probability of eternal inflation occurring for two values of $m_{0}$. Recalling the definition of $m_{0}$ from Eq. \eqref{eq:newdefs}, we see that $m_{0} = \frac{1}{\sqrt{3}}$, 
for $m = 10^{12} \mbox{ GeV}$, and
$m_{0} = \frac{10}{\sqrt{3}}$,
for $m = 10^{13} \mbox{ GeV}$.  

First, for $m_{0} = \frac{1}{\sqrt{3}}$, we calculate
\begin{eqnarray}
P\left(X\right) &=& \nonumber \\
&=& \int_{-1}^{-\sqrt{2/3}}0.253641 e^{-\frac{1}{2} X^2 \left(3 X^2-\frac{16}{3}\right)} dX + \int_{\sqrt{2/3}}^{1} 0.253641 e^{-\frac{1}{2} X^2 \left(3 X^2-\frac{16}{3}\right)} dX \nonumber \\
&=& 0.299.
\end{eqnarray}

Second, for $m_{0} = \frac{10}{\sqrt{3}}$, we calculate
\begin{eqnarray}
P(X) &=& \nonumber \\
&=& \int_{\sqrt{2/3}}^{1} 3.11109 e^{-\frac{1}{2} X^2 \left(3 X^2+\frac{182}{3}\right)} dX \nonumber \\
&=& 4.87 \times 10^{-11}.
\end{eqnarray}

The BICEP2 experiment recently determined that $H \approx 10^{14} \mbox{ GeV}$ ($10^{-5}$ in natural units) during inflation \cite{2014PhLB..734...21H}. Using this value, we now try calculating the eternal inflation probability with $m_{0} = \frac{1}{100 \sqrt{3}}$ and $m_{0} = \frac{1}{10 \sqrt{3}}$. We obtain that for  $m_{0} = \frac{1}{100 \sqrt{3}}$, 
\begin{eqnarray}
P\left(X\right) &=& \nonumber \\
&=& \int_{-1}^{-\sqrt{2/3}}0.21677 e^{-\frac{1}{2} X^2 \left(3 X^2-\frac{89999}{15000}\right)} dX + \int_{\sqrt{2/3}}^{1} 0.21677 e^{-\frac{1}{2} X^2 \left(3 X^2-\frac{89999}{15000}\right)} dX \nonumber \\
&=& 0.3367.
\end{eqnarray}

While for $m_{0} = \frac{1}{10 \sqrt{3}}$, we see that
\begin{eqnarray}
P\left(X\right) &=& \nonumber \\
&=& \int_{-1}^{-\sqrt{2/3}}0.217118 e^{-\frac{1}{2} X^2 \left(3 X^2-\frac{899}{150}\right)} dX + \int_{\sqrt{2/3}}^{1} 0.217118 e^{-\frac{1}{2} X^2 \left(3 X^2-\frac{899}{150}\right)} dX \nonumber \\
&=& 0.3363.
\end{eqnarray}

We thus see that the highest probability of eternal inflation occurring, about 33.67\% occurs for $m_{0} = 1/(100\sqrt{3})$, which corresponds to $m = 10^{-7}$, and $H = 10^{-5}$.  The second-highest probability of eternal inflation occurring was 33.63\% for $m_{0} = 1/(10\sqrt{3})$, which corresponds to $m = 10^{-6}$ and $H = 10^{-5}$. We also see that the probability of eternal inflation is infinitesimally small when $m_{0} = 10/\sqrt{3}$, which corresponds to $m = 10^{-6}$ and $H = 10^{-7}$.

\section{Conclusions}
In this paper, we studied the phenomenon of stochastic eternal inflation for a chaotic inflation potential in a Bianchi Type I spacetime background. After deriving the appropriate stochastic Klein-Gordon equation, we derived the conditions for eternal inflation to occur. We showed that for eternal inflation to occur, the amount of anisotropy had to be very small. In fact, we showed that eternal inflation would only take place if the shear anisotropy variables took on values within a small region of the interior of the Kasner circle. We  then  calculated the probability of eternal inflation occurring based on techniques from stochastic calculus. 

We found that the highest probability of eternal inflation occurring, about 33.67\% occured for $m_{0} = 1/(100\sqrt{3})$, which corresponds to $m = 10^{-7}$, and $H = 10^{-5}$.  The second-highest probability of eternal inflation occurring was 33.63\% for $m_{0} = 1/(10\sqrt{3})$, which corresponds to $m = 10^{-6}$ and $H = 10^{-5}$. We saw that the probability of eternal inflation was infinitesimally small when $m_{0} = 10/\sqrt{3}$, which corresponds to $m = 10^{-6}$ and $H = 10^{-7}$.

\section{Acknowledgements}
This research was partially supported by a grant given to MCH from the Natural Sciences and Engineering Research Council of Canada. In addition, ISK would like to thank Sigbj{\o}rn Hervik, Oleksandr Pavlyk, and Alexey Kuznetsov for helpful comments and suggestions. We would also like to thank the anonymous referee for helpful comments and suggestions.

\newpage 
\bibliographystyle{ieeetr} 
\bibliography{sources}

\end{document}